# High-Efficiency Graphene Oxide Moiré Lenses Enabled by Cyclic Self-Assembly and Laser Nanoprinting

Guiyuan Cao[1,2,#], Xinyuan Du[1,#], Han Lin[2], Baohua Jia[2,*], Shibiao Wei[1,*]

[1]*Nanophotonics Research Center, Institute of Microscale Optoelectronics and State Key Laboratory of Radio Frequency Heterogeneous, Shenzhen University, Shenzhen 518060, China*

[2]*Centre for Atomaterials and Nanomanufacturing (CAN), School of Science, RMIT University, Melbourne, 3000 VIC, Australia*

Email: *baohua.jia@rmit.edu.au; weishibiao@szu.edu.cn*

**Abstract** Compact optical systems with continuously tunable focal lengths are highly desirable for machine vision, biomedical imaging, and miniaturized cameras. Rotational Moiré lenses offer a planar approach to focal-length tuning, but metasurface-based implementations typically require complex nanofabrication processes that complicate large-area production. Here, we present a cyclic, mask-free fabrication strategy that combines layer-by-layer self-assembly of two-dimensional graphene oxide (GO) and poly(diallyldimethylammonium chloride) (PDDA) with femtosecond laser nanoprinting. Repeated GO-PDDA assembly and spatially selective ablation enable atomic-scale thickness control and the construction of precise multilevel phase profiles, reducing phase-quantization errors and improving focusing efficiency. The measured focusing efficiencies of individual GO lenses increase from 22.4% for two phase levels to 54.8% for six levels. Using this six-level modulation strategy, the fabricated GO Moiré lens achieves continuous focal-length tuning from approximately 10 to 90 mm. Full-colour imaging with magnifications ranging from 1× to 9× is experimentally demonstrated. These results establish a cyclic, mask-free fabrication strategy for efficient multilevel GO flat optics and demonstrate its potential for compact, reconfigurable imaging.

## Introduction

Compact optical systems with continuously tunable focal lengths are highly desirable for applications in machine vision, biomedical imaging, wearable devices, autonomous sensing, and miniaturized cameras[1-4]. Conventional zoom systems generally rely on the axial displacement of multiple refractive lenses, resulting in bulky configurations and complex mechanical assemblies. To overcome these limitations, various tunable optical components have been developed, including electrically tunable liquid-crystal lenses, mechanically stretchable lenses, and thermally or optically reconfigurable metasurfaces. These approaches offer compact optical architectures, with focal-length tuning achieved through material deformation or externally driven changes in the optical properties of the constituent materials[5-8].

Rotational Moiré lenses provide an attractive approach to focal-length tuning in compact optical systems [9-11]. A typical Moiré lens consists of two closely spaced phase elements with complementary angular phase distributions. Relative rotation between the two elements produces an effective quadratic phase profile, enabling continuous focal-length tuning without axial translation. This distinctive mechanism offers a planar architecture, simple rotational actuation, and a potentially large tuning range. Recent advances in rotational Moiré optics have employed dielectric metasurfaces and liquid-crystal-based phase elements to demonstrate dynamic focusing, zoom imaging, and optical beam manipulation [12-15]. In particular, dielectric metasurfaces enable precise wavefront control and high-quality focusing through engineered subwavelength nanostructures. However, fabricating these nanostructures generally requires high-resolution lithography, material deposition, and precise etching, which complicates scalable and cost-effective production, particularly for large-area optical components.

In this context, two-dimensional (2D) materials provide an attractive platform for advancing ultrathin planar optics towards atomic- and close-to-atomic-scale manufacturing. Their atomically thin building blocks, compatibility with large-area preparation, and direct patternability offer a unique pathway for translating atomic-scale material control into scalable optical-device fabrication [16-19]. Among them, graphene oxide (GO) is particularly attractive for planar optics because it combines laser-tailorable optical properties with layer-by-layer thickness control[20-25]. Focused-laser processing can locally modify the refractive index, absorption, and thickness of GO through photoreduction or material ablation, enabling the direct patterning of optical phase and amplitude without resist-based lithography, physical masks, or complex multistep cleanroom processing. This approach has been employed to fabricate ultrathin lenses, orbital-angular-momentum generators, and other planar optical components [26-31]. Moreover, layer-by-layer self-assembly (SA) enables atomic-scale control over film thickness across large areas, providing a bottom-up manufacturing route that translates atomically derived 2D building blocks into macroscale functional optical devices [32-34]. This combination of atomic-scale thickness programmability, large-area assembly, and direct laser patterning makes GO particularly suitable for the scalable and cost-effective manufacturing of integrated planar optics.

Despite these advantages, the practical imaging performance of existing GO metalenses remains limited by their relatively low focusing efficiency [21-23, 34]. This limitation arises from several factors, including the approximately binary modulation produced by conventional single-step laser patterning, the intrinsic absorption of reduced GO, and incomplete phase coverage. In a quasi-binary GO lens, as illustrated in Fig. 1a, a considerable portion of the incident optical energy is distributed into undesired diffraction orders and background fields rather than being directed towards the target focus. This problem becomes more pronounced in a rotational Moiré system, in which two patterned elements jointly produce the effective focusing wavefront. Uncontrolled diffraction channels and imperfect interference between the two elements can substantially reduce the energy delivered to the desired tunable focal spot, thereby limiting image brightness, contrast, and practical applicability.

Multilevel wavefront modulation provides a promising route to reduce phase-quantization losses and improve focusing efficiency [35, 36]. In this approach, the continuous phase profile is discretized into a series of phase levels and approximated by step-like zones, as shown in Fig. 1b. Adjacent levels have an equal height increment *h*, corresponding to a uniform phase increment, while their lateral widths *w* vary according to the designed phase distribution, enabling a more accurate approximation of the ideal continuous wavefront. Consequently, phase-quantization errors and unwanted higher-order diffraction are reduced, directing more optical energy into the desired diffraction order. For an ideal lossless multilevel phase lens operating at the design wavelength, increasing the number of phase levels from two to six raises the theoretical first-order diffraction efficiency from 40.5% to 91.2% [37, 38], as illustrated in Fig. 1a. Importantly, introducing multiple modulation levels into a Moiré lens not only enhances the efficiency of each individual diffractive element, but also enables a more accurate realization of the two complementary phase profiles, thereby providing improved control over the distribution of optical energy among the diffraction orders of the combined Moiré system. By reducing phase-quantization errors, multilevel modulation directs more optical power into the desired tunable focusing order while suppressing parasitic diffraction orders and background. Previous studies have demonstrated multilevel GO micro-optics by tuning laser-induced photoreduction and photoablation. However, the coupled changes in refractive index, absorption, and film thickness make independent control of phase and amplitude difficult and therefore do not necessarily translate into improved focusing efficiency [39, 40]. Therefore, a fabrication strategy that defines the modulation levels through precise film-thickness control, rather than through different degrees of laser-induced photoreduction, is highly desirable.

Here, we present a high-efficiency multilevel GO Moiré metalens fabricated by cyclic layer-by-layer SA and femtosecond laser direct nanoprinting. As shown in Fig. 1b, each fabrication cycle consists of depositing a prescribed number of GO layers, followed by selective laser ablation of designated regions. Repeating this deposition–ablation cycle progressively constructs the multilevel optical modulation profiles of the Moiré elements. By combining precise thickness control with spatially selective laser ablation, this approach enables multilevel wavefront modulation through a simple, mask-free, and scalable fabrication process. The resulting multilevel profiles more faithfully reproduce the designed wavefronts, thereby reducing phase-quantization errors and improving optical efficiency. As illustrated in Fig. 1c, relative rotation between the two complementary multilevel GO Moiré elements enables continuous focal-length tuning, thereby supporting zoom imaging in an ultrathin planar architecture.

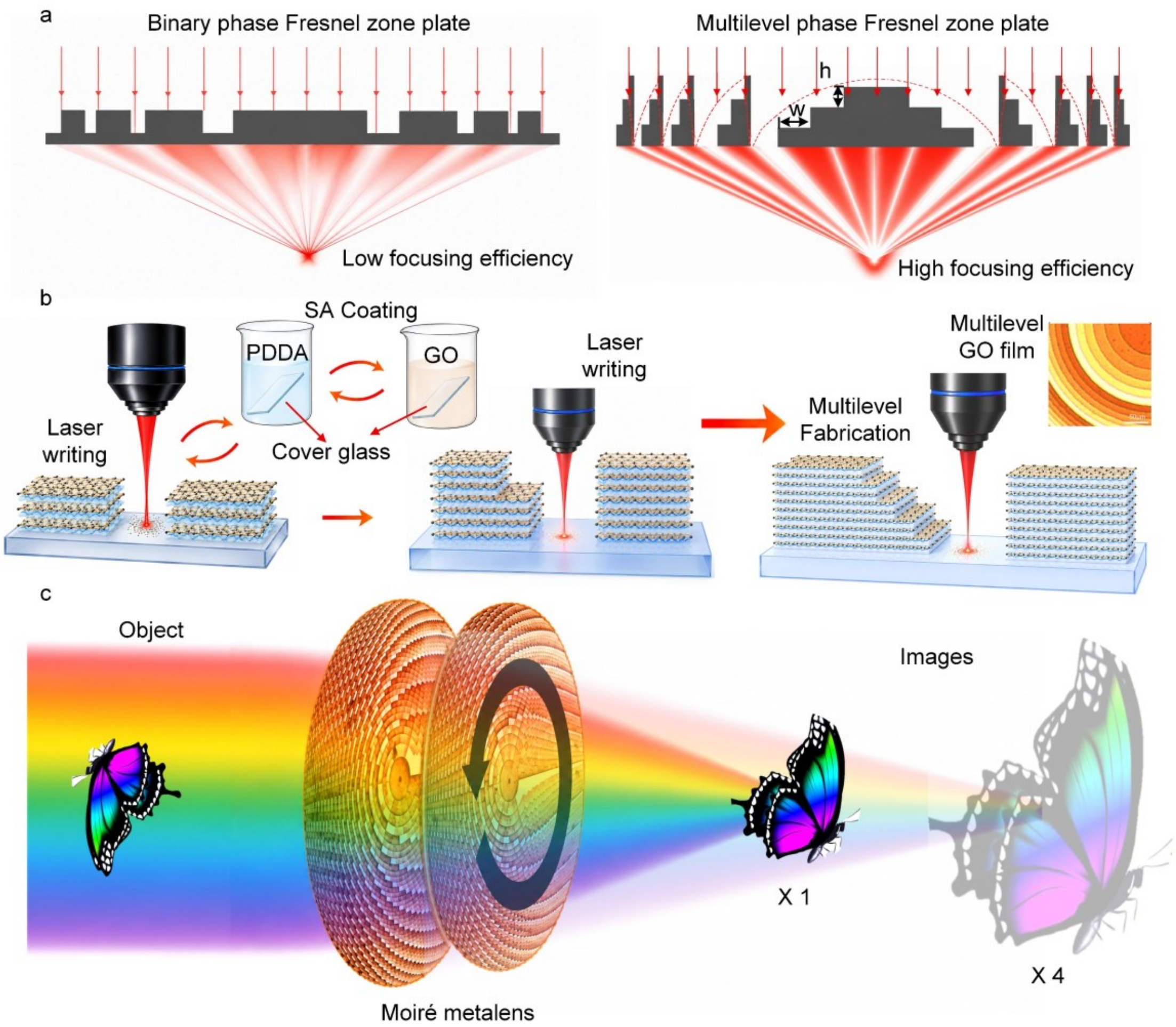

Fig. 1. Multilevel GO Moiré lens for variable-magnification imaging. (a) Schematic illustration of enhanced focusing efficiency through multilevel phase modulation compared with a conventional binary Fresnel zone plate. (b) Fabrication of multilevel phase profiles through cyclic layer-by-layer SA and femtosecond-laser direct nanoprinting. In each cycle, a GO film with a prescribed thickness is deposited by SA, followed by selective femtosecond-laser ablation of designated regions. Repeating the deposition–ablation cycle progressively constructs the designed multilevel phase profile. (c) Schematic of variable-magnification imaging enabled by the GO Moiré lens. Relative rotation between the two complementary elements continuously tunes the effective focal length, enabling images with different magnifications to be obtained.

Our experimental results show that increasing the number of phase levels from two to four and six progressively suppresses parasitic diffraction and improves the focusing efficiency from 22.4% to 36.3% and 54.8%, respectively. These measured efficiencies are lower than the ideal lossless predictions primarily because of optical absorption in the GO films, which reduces the optical power available for focusing. Nevertheless, the progressive improvement confirms that finer phase discretization enables more effective utilization of the incident optical energy. Based on the six-level modulation strategy, the fabricated GO Moiré lens provides continuous focal-length tuning from approximately 10 to 90 mm through relative in-plane rotation of its two complementary elements. Under white-light LED illumination, the device further demonstrates broadband full-colour variable-magnification imaging, with representative magnifications ranging from 1× to 9×. These results demonstrate that multilevel wavefront modulation can simultaneously enhance optical efficiency and support practical broadband imaging in a rotationally reconfigurable GO platform.

This work advances GO Moiré lenses towards practical high-efficiency, high-contrast and broadband variable-magnification imaging in an ultrathin planar architecture, with potential applications in portable microscopy, machine vision, and integrated imaging systems. More broadly, the proposed cyclic fabrication strategy combines atomic-scale thickness control through large-area layer-by-layer SA with spatially selective, mask-free femtosecond-laser patterning, providing a scalable manufacturing route for complex multilevel wavefront profiles and efficient, reconfigurable planar optics.

## 2. Precise fabrication of multilevel graphene oxide lenses

Multilevel GO lenses require precise control over the optical phase modulation introduced by the GO film. In this work, a GO film consisting of alternating GO and poly(diallyldimethylammonium chloride) (PDDA) layers was used. It is fabricated through a layer-by-layer SA process, as illustrated by the coating cycle in Fig. 1b. Further details are provided in Supplementary Section 1 and Fig. S1. Owing to the presence of abundant oxygen-containing functional groups, GO is negatively charged in aqueous solution, whereas PDDA is positively charged. The electrostatic attraction between these oppositely charged species enables their sequential deposition, thereby forming a uniform GO–PDDA multilayer film. Each PDDA–GO bilayer has a thickness of approximately 3 nm, while the resulting multilayer film exhibits a typical surface roughness of less than 2 nm. The optical phase modulation introduced by the film can therefore be precisely controlled by varying the number of deposited bilayers. To enable the accurate construction of discrete phase levels, it is essential to establish the quantitative relationship between the number of bilayers and the resulting phase difference.

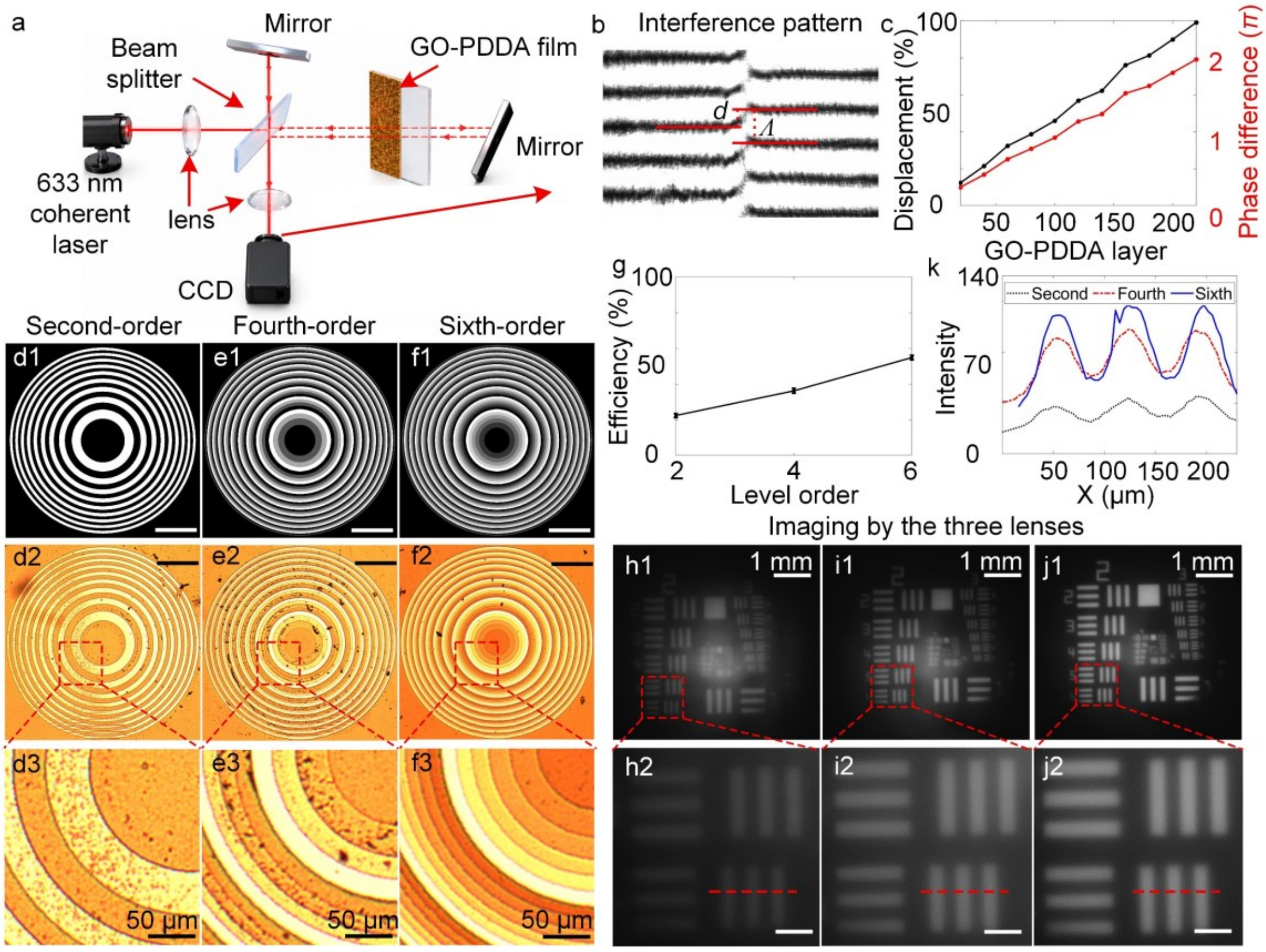

Fig. 2. Precise fabrication and optical characterization of multilevel GO lenses with high focusing efficiency and high-quality imaging capability. (a) Schematic of the interference-based system used to calibrate the relationship between the number of deposited GO–PDDA bilayers and the induced phase modulation. (b) Interference fringes obtained from the GO-film-coated and bare-substrate regions, where *Δd* denotes the relative fringe displacement, and *Λ* is the fringe period. (c) Measured fringe displacement and corresponding phase difference as functions of the number of deposited GO–PDDA bilayers. (d1-f1) Designed lens patterns discretized into two, four, and six phase levels, respectively. (d2-f2) Fabricated GO lenses with two, four, and six phase levels, respectively. (d3-f3) Enlarged views of the regions indicated by the red dashed rectangles in (d2–f2). (g) Measured focusing efficiencies of the two-, four-, and six-level GO lenses under 633 nm illumination. (h1-j1) Images of a 1951 USAF resolution target captured using the two-, four-, and six-level GO lenses, respectively, under illumination from a 633 nm LED. (h2-j2) Enlarged views of the regions indicated by the red dashed rectangles in (h1–j1), showing elements 5 and 6 of Group 2. (k) Intensity profiles extracted along the red dashed lines in (h2–j2). Scale bars: 200 μm.

An interference-based measurement system was constructed using a 633 nm coherent laser to experimentally characterize this relationship, as shown in Fig. 2a. The incident beam was divided by a beam splitter (BS) into a reference beam and a sample beam. The reference beam was reflected by a mirror and directed toward the CCD camera, whereas the sample beam passed through the GO–PDDA-coated and uncoated regions of the cover glass. The two beams were subsequently recombined to generate interference patterns corresponding to the film-coated and bare-substrate regions, as shown in Fig. 2b. The additional optical path length introduced by the GO–PDDA film produced a relative displacement of the interference fringes between the two regions. The resulting phase difference was calculated from the measured fringe displacement according to $\Delta\varphi = 2\pi\Delta d/\Lambda$, where $\Delta\varphi$ is the phase difference between the coated and uncoated regions, $\Delta d$ is the relative displacement of the interference fringes, and $\Lambda$ is the fringe period. The relationship between the number of deposited bilayers and the induced phase modulation was then experimentally determined by measuring the relative fringe displacement of samples with different bilayer numbers. As shown in Fig. 2c, both the measured fringe displacement and the corresponding phase difference increase approximately linearly with the number of deposited bilayers. Each GO–PDDA bilayer introduces a phase shift of approximately 0.009π, resulting in a total phase modulation of approximately 2π for 220 bilayers. This experimentally established calibration provides the basis for accurately determining the required number of bilayers and constructing the discrete phase levels of the multilevel GO lenses.

Based on the conventional hyperbolic phase distribution, $\varphi(r) = \frac{2\pi}{\lambda}(\sqrt{r^2 + f^2} - f)$, where $r$ is the radial coordinate, $f$ is the designed focal length, and $\lambda$ is the design wavelength, a lens phase profile with radius of 500 μm and a focal length of 20 mm was design at 633 nm. The continuous phase profile was then discretized into two, four, and six phase levels. For an $m$-level lens, the discrete phase values are separated by $2\pi/m$, giving a maximum phase modulation of $2(m\text{-}1)\pi/m$. The resulting two-, four-, and six-level lens patterns are shown in Figs. 2d1–f1, respectively. Based on the experimentally calibrated relationship between the number of GO–PDDA bilayers and the induced phase difference, the maximum numbers of bilayers required for the two-,

four-, and six-level GO lenses were determined to be 110, 165, and 183 bilayers, corresponding to maximum phase modulations of $\pi$, $3\pi/2$, and $5\pi/3$. The lenses were subsequently fabricated through cyclic layer-by-layer deposition and spatially selective laser ablation, with the detailed fabrication procedure provided in Supplementary Fig. S2. The fabricated lenses are shown in Figs.2 d2-f2, while Figs. 2d3-f3 present magnified views of the regions marked by the red rectangles in Figs. 2 d2-f2. Under an optical microscope, the different phase levels exhibit distinct colors owing to variations in the GO-film thickness, allowing the multilevel structures to be clearly distinguished.

The focusing efficiencies of the three GO lenses were measured under 633 nm illumination, as shown in Fig. 2g. The focusing efficiency was defined as the ratio of the optical power contained within the focal spot to the optical power incident on the lens. The measured efficiencies were 22.4%, 36.3%, and 54.8% for the two-, four-, and six-level GO lenses, respectively. The focusing efficiency therefore increased consistently with the number of phase levels, confirming that finer phase discretization directs a larger fraction of the incident optical power into the desired focus. Nevertheless, owing primarily to optical absorption in the GO film, the measured efficiency of the six-level lens remained below the theoretical value predicted for an ideal lossless multilevel phase lens.

The imaging performance of the three GO lenses was evaluated using a 1951 USAF resolution target illuminated by a 633 nm LED, as shown in Figs. 2h1, i1, and j1. As the number of phase levels increased, the images became progressively brighter, and the target features exhibited improved visibility. Enlarged images of elements 5 and 6 in Group 2 are presented in Figs. 2h2, i2, and j2, where the corresponding line pairs can be clearly resolved by all three lenses. In particular, element 6 corresponds to a spatial frequency of 7.13 line pairs/mm, equivalent to a bar width of approximately 70 μm. To quantitatively compare their imaging performance, intensity profiles along the red dashed lines in Figs. 2h2, i2, and j2 were extracted and plotted in Fig. 2k. The peak-to-valley intensity differences increased from 16 for the two-level lens to 33 and 58 for the four- and six-level lenses, respectively. Under the same measurement conditions, the higher image intensity and stronger absolute intensity modulation make the resolved features less susceptible to imaging noise and improve their visibility. Together with the measured focusing efficiencies, these results demonstrate that multilevel phase modulation substantially improves both the focusing efficiency and practical imaging performance of the GO lenses.

## 3. Design and characterization of GO Moiré lenses

Building on the fundamental investigation of multilevel GO lenses, we extended the multilevel wavefront-modulation strategy to rotational Moiré lenses. The phase profiles of the two complementary Moiré elements $T_1$ and $T_2$ were designed using the following complex transmission functions:

$$T_1(r,\varphi) = \exp(iar^2\varphi),$$

$$T_2(r,\varphi) = \exp(-iar^2\varphi),$$

where $a$ is the phase-modulation coefficient, $r$ is the radial coordinate, and $\varphi$ is the azimuthal angle. When a relative rotation angle $\theta$ is introduced between elements $T_1$ and $T_2$, their combined transmission function becomes

$$T_{total} = \exp(iar^2\varphi)\exp[-iar^2(\varphi-\theta)] = \exp(ia\theta r^2).$$

The corresponding effective focal length is given by

$$f = \frac{\pi}{a\theta\lambda}.$$

For a fixed operating wavelength $\lambda$ and phase-modulation coefficient $a$, the effective focal length $f$ is inversely proportional to the relative rotation angle $\theta$. Continuous focal-length tuning can therefore be achieved by rotating the two complementary elements relative to each other.

To implement the continuously varying phase profiles using multilevel GO structures, the phase distributions of both Moiré elements were discretized using an upward-rounding scheme. This quantization rule preserves a consistent phase-level assignment across the phase-wrapping boundary and avoids an abrupt change in the discretized profile at $\varphi = \pi$, which would otherwise divide the lens into sector-shaped regions with inconsistent local phase gradients. Figs. 3a1 and 3b1 show the six-level discretized phase distributions of the $T_1$ and $T_2$ elements, respectively. The resulting profiles exhibit stepwise curved-ring patterns whose radial spacing decreases nonlinearly with increasing $r$, leading to a progressively higher feature density toward the outer regions of the elements.

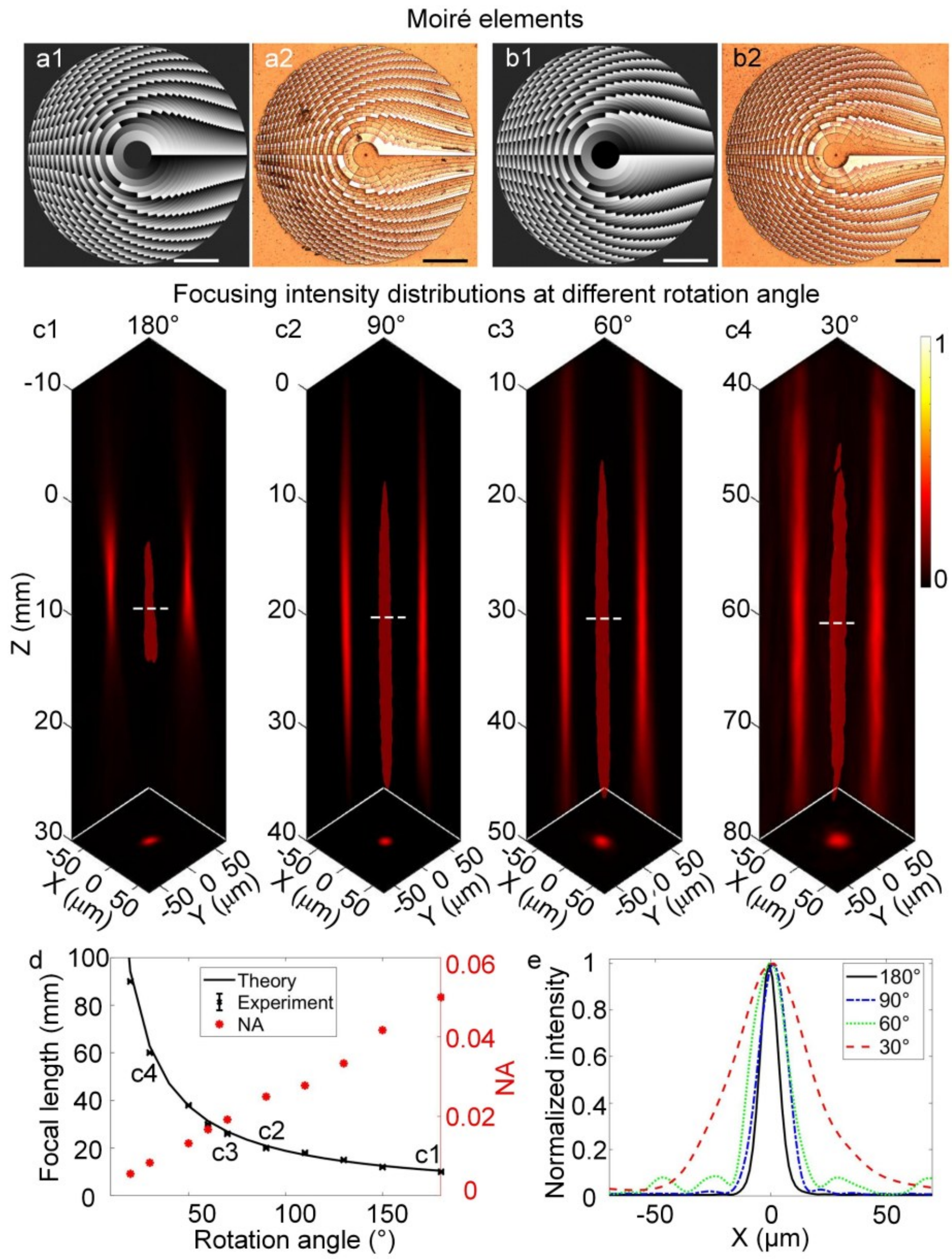

Fig. 3. Fabrication and optical characterization of the six-level GO Moiré lens. (a1-b1) Six-level discretized phase distributions of the complementary $T_1$ and $T_2$ elements, respectively. (a2-b2) Fabricated six-level GO Moiré lens elements corresponding to $T_1$ and $T_2$, respectively. (c1-c4) Representative 3D focal-intensity distributions and the corresponding longitudinal intensity distributions in the *x*-*z* and *y*-*z* planes and transverse intensity distributions in the *x*-*y* focal plane at relative rotation angles of 180°, 90°, 60°, and 30°, respectively. (d) Experimentally measured and simulated focal lengths and the corresponding NA at different rotation angles. (e) Experimentally measured and simulated lateral FWHMs of the focal spots at relative rotation angles of 30°, 60°, 90°, and 180°. The bar scale is 200 μm.

Following the phase discretization, the complementary six-level GO Moiré lens elements were fabricated using the cyclic layer-by-layer SA and femtosecond-laser ablation process described above, as shown in Figs. 3a2 and 3b2. The six discrete GO thickness levels can be readily distinguished by their different optical contrasts and colours, providing qualitative verification of the successful construction of the designed multilevel profiles. These results further demonstrate that the cyclic fabrication strategy can be extended from conventional multilevel lenses to more complex rotational Moiré optical elements.

At an incident wavelength of 633 nm, the focal length of the GO Moiré lens was experimentally tuned from approximately 10 to 90 mm by varying the relative rotation angle $\theta$ between the two elements. The characterization setup is illustrated in Supplementary Fig. S3. Four representative focusing performance at relative rotation angles of 180°, 90°, 60°, and 30°, corresponding to focal lengths of approximately 10, 20, 30, and 60 mm, respectively, were characterized. The corresponding three-dimensional (3D) intensity distributions, together with the longitudinal intensity profiles in the *x-z* and *y-z* planes, and the transverse intensity profiles in the *x-y* focal plane are shown in Figs. 3c1- c4. Well-defined focal spots were obtained at all four rotation angles. Additional measured and simulated focal lengths and the corresponding numerical aperture (NA) at different rotation angles are plotted in Fig. 3d. The focal lengths corresponding to Figs. 3c1-c4 are approximately 10, 20, 30, 60, and the NA is 0.05, 0.025, 0.017, 0.008, separately. The experimental results agree well with the simulations, confirming the theoretically predicted relationship between the relative rotation angle and the focal length. Continuous focal-length tuning was therefore experimentally demonstrated by rotating the two elements relative to each other. Additional data are provided in Supplementary Table 1. It was further observed that the axial full width at half maximum (FWHM) of the focal spot along the z-direction increases with increasing the focal length. This broadening results from the decrease in the effective NA at longer focal lengths. A similar trend is observed for the transverse focal plane, as shown in Fig. 3e. At relative rotation angles of 180°, 90°, 60°, and 30°, the measured FWHMs along the x-direction are 41.34, 20.75, 13.89, and 9.43 μm. The progressive decrease in the lateral focal-spot size agrees well with the simulated results and is consistent with the reduction in the effective NA.

## 4. Broadband and zoom imaging of GO Moiré lens

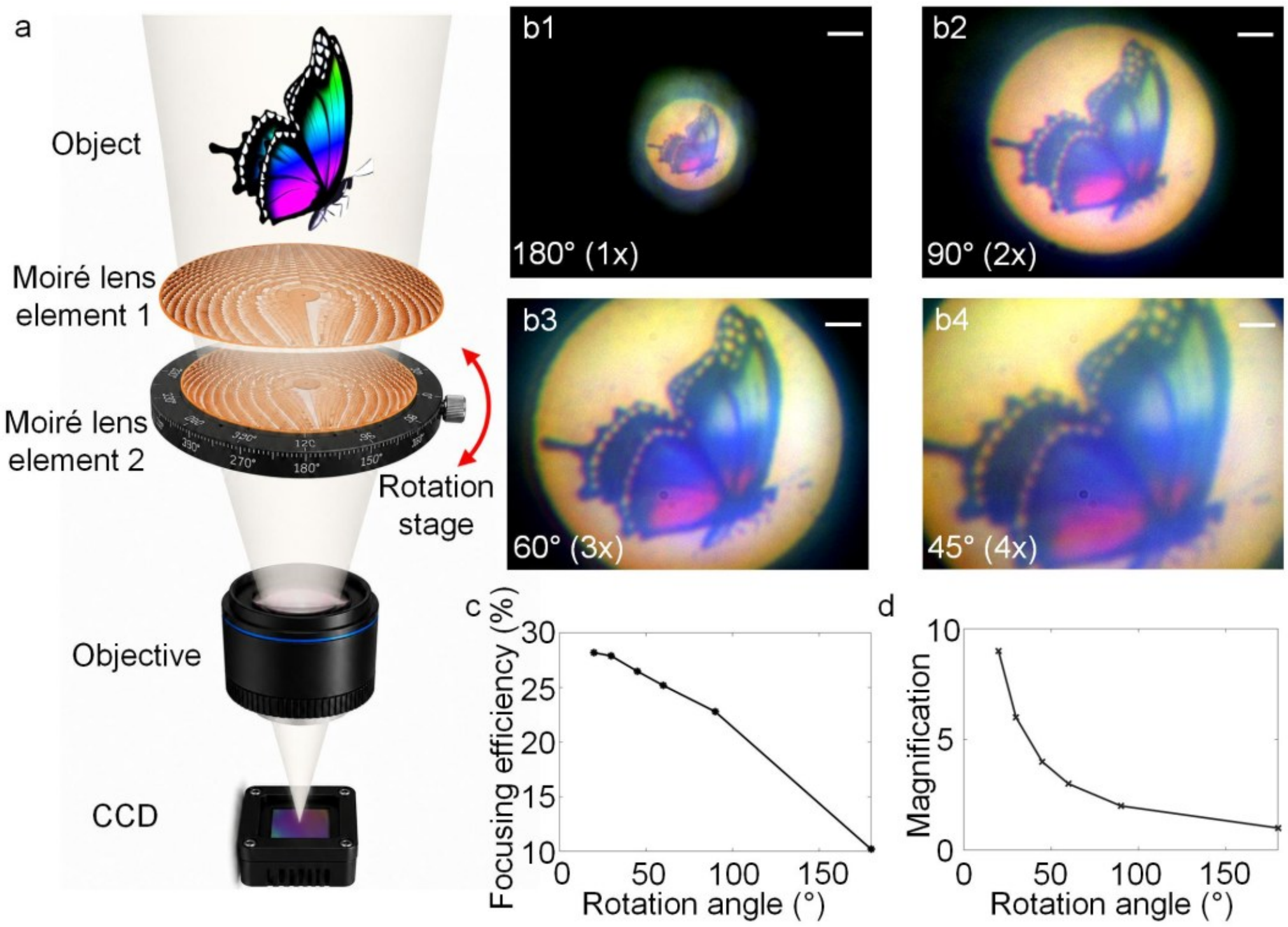

Fig. 4. Broadband zoom-imaging capability of the varifocal GO Moiré lens. (a) Schematic of the variable-magnification imaging system based on the GO Moiré lens. (b1-b4) Experimentally captured representative full-colour images under white-light LED illumination at relative rotation angles of 180°, 90°, 60°, and 45°, with magnifications of approximately 1× to 2×, 3×, and 4×, respectively. (c) Measured focusing efficiency of the GO Moiré lens at selected relative rotation angle. (d) Representative imaging magnifications achieved at selected relative rotation angles. The bar scale is 2 mm.

The broadband zoom-imaging capability of the multilevel GO Moiré lens was experimentally demonstrated using a full-colour butterfly pattern illuminated by a white-light LED. The experimental configuration is illustrated in Fig. 4a. The two complementary GO Moiré lens elements were coaxially mounted, with one element fixed and the other attached to a rotation stage. The image formed by the combined Moiré lens was collected by an objective and recorded using a CCD camera. Relative in-plane rotation between the two elements varied the effective focal length of the Moiré lens, enabling zoom imaging without axial translation of the optical components.

Full-colour imaging with magnifications ranging from approximately 1× to 9× was experimentally demonstrated under white-light LED illumination. Figs. 4b1–b4 show representative images captured at relative rotation angles of 180°, 90°, 60°, and 45°, corresponding to focal lengths of approximately 10, 20, 30, and 38 mm, respectively. By adjusting the object and image distances for each focal length, magnifications of approximately 1×, 2×, 3×, and 4× were obtained, respectively. The images became progressively enlarged, while the multicolour features of the object remained clearly distinguishable at all four magnifications under white-light illumination, demonstrating the broadband imaging capability of the GO Moiré lens. The experimentally measured focusing efficiencies range from approximately 10% to 30% at an incident wavelength of 633 nm. These efficiencies and the imaging magnifications at various relative rotation angles are summarized in Figs. 4c and 4d, respectively. The corresponding numerical data are listed in Supplementary Table 2. Although these measurements represent discrete imaging states, intermediate magnifications can be accessed through continuous focal-length tuning and corresponding adjustment of the imaging configuration. Together with the enhanced focusing efficiency and continuous rotation-controlled focal-length tuning demonstrated above, these broadband full-colour imaging results show that the multilevel GO Moiré lens can translate tunable wavefront control into practical variable-magnification imaging. This combination of high optical efficiency, continuous focal-length tunability, and broadband imaging capability establishes the multilevel GO Moiré lens as a promising platform for compact and reconfigurable planar imaging systems.

## 5. Conclusion

In conclusion, cyclic GO–PDDA layer-by-layer SA combined with selective femtosecond-laser ablation provides a controlled, mask-free approach to constructing multilevel GO optical elements. The calibrated relationship between bilayer number and optical phase allows prescribed phase profiles to be built incrementally. Under 633 nm illumination, the measured focusing efficiencies of individual two-, four-, and six-level GO lenses were 22.4%, 36.3%, and 54.8%, respectively. This progression demonstrates the benefit of finer phase discretization for directing incident optical energy into the desired focus. The extension to complementary six-level Moiré elements further demonstrates the applicability of this fabrication approach to more complex phase profiles and reconfigurable optical functions. Relative rotation enabled continuous focal-length tuning from approximately 10 to 90 mm, while full-colour imaging under white-light LED illumination was demonstrated at magnifications ranging from 1× to 9×. This work translates the thickness programmability of assembled 2D materials into functional optical devices. The combination of layer-by-layer assembly and mask-free laser patterning offers a flexible route to multilevel flat optics beyond conventional lens profiles. Refining phase control could further improve focusing efficiency, while integration with compact rotational actuation could facilitate practical imaging applications. These findings provide a basis for developing efficient, reconfigurable optical components from two-dimensional materials.

## Author Contributions

[#]These authors contributed equally: Guiyuan Cao and Xinyuan Du

## Acknowledgments

**Funding:** This work was supported by the National Natural Science Foundation of China (62205214 and 62175162); Shenzhen University 2035 Program for Excellent Research (2022C013); The Australian Research Council through the Discovery Project scheme (Grant No. DP190103186, FT210100806, and FT220100559), the Linkage Project scheme (LP210200345, LP210100467), the Industry Early Career Fellowship (IE250100125), and the Industrial Transformation Training Centres scheme (IC180100005), and the Centre of Excellence Program (CE230100006). The Science and Technology Innovation Commission of Shenzhen (JCYJ20241202124428038).

## Disclosures

The authors declare no conflicts of interest.

## Data Availability

Data underlying the results presented in this paper are not publicly available at this time but may be obtained from the authors upon reasonable request.